\documentclass[conference]{IEEEtran}
\usepackage{cite}
\usepackage{amsmath,amssymb,amsfonts}
\usepackage{algorithmic}
\usepackage{graphicx}
\usepackage{textcomp}
\usepackage{xcolor}
\usepackage[hyphens]{url}
\usepackage[capitalise,noabbrev]{cleveref}
\usepackage{array}
\usepackage{color}
\usepackage{colortbl}
\usepackage{booktabs}
\usepackage{listings}
\usepackage{balance}
\usepackage{multirow,makecell}

\def\BibTeX{{\rm B\kern-.05em{\sc i\kern-.025em b}\kern-.08em
    T\kern-.1667em\lower.7ex\hbox{E}\kern-.125emX}}

\title{The Intel\textsuperscript{\textregistered}~Programmable and Integrated \\ Unified Memory Architecture (PIUMA) \\ Graph Analytics Processor}
\author{Sriram Aananthakrishnan ~~ Shamsul Abedin ~~ Vincent Cav\'{e} ~~ Fabio Checconi ~~ Kristof Du Bois \\
Stijn Eyerman ~~ Joshua B.~Fryman ~~ Wim Heirman ~~ Jason Howard ~~ Ibrahim Hur ~~ Samkit Jain \\
Marek M.~Landowski ~~ Kevin Ma ~~ Jarrod Nelson ~~ Robert Pawlowski ~~ Fabrizio Petrini \\
Sebastian Szkoda ~~ Sanjaya Tayal ~~ Jesmin Jahan Tithi ~~ Yves Vandriessche \vspace{3mm} \\
{\em Intel Corporation}}

\begin{document}
\maketitle
\thispagestyle{plain}
\pagestyle{plain}


\newcommand{\ignore}[1]{}
\newcommand{\warn}[1]{\textcolor{red}{#1}}
\newcommand{\todo}[1]{{ \bf \em [TODO: #1]}}
\newcommand{\eg}{{\em e.g.,\ }}
\newcommand{\ie}{{\em i.e.,\ }}
\newcommand{\etal}{{\em et al.}}
\newcommand\PUMAreg{Intel\textsuperscript{\textregistered}~PIUMA}
\newcommand\Intel{Intel\textsuperscript{\textregistered}~}

\begin{abstract}

High performance large scale graph analytics are essential to timely analyze relationships in big data sets.
Conventional processor architectures suffer from inefficient resource usage and bad scaling on those workloads.
To enable efficient and scalable graph analysis, Intel\textsuperscript{\textregistered}~developed the
Programmable Integrated Unified Memory Architecture (PIUMA) as a part of the DARPA Hierarchical Identify Verify Exploit (HIVE) program.
PIUMA consists of many multi-threaded cores,
fine-grained memory and network accesses, a globally shared address space, powerful offload engines and a tightly integrated optical interconnection network.
By utilizing co-packaged optical silicon photonics and extending the on-chip mesh protocol directly to the optical fabric,
all PIUMA chips in a system are glued together in a large virtual die which allows for extremely low socket-to-socket latencies
even as the system scales to thousands of sockets.
Performance estimations project that a PIUMA node will outperform a conventional compute node by one to two orders of magnitude. 
Furthermore, PIUMA continues to scale across multiple nodes, which is a challenge in conventional multi-node setups.

This paper presents the PIUMA architecture, and documents our experience in designing and building a prototype chip
and its bring-up process. We summarize the methodology for our
co-design of the architecture together with the software stack using simulation tools and FPGA emulation. These tools provided early
performance estimations of realistic applications and allowed us to implement many optimizations across the hardware,
compilers, libraries and applications.
We built the PIUMA chip as a 316mm$^2$ 7nm FinFET CMOS die and constructed a 16-node system.
PIUMA silicon has successfully powered on demonstrating key aspects of the architecture,
some of which will be incorporated into future Intel products.
\end{abstract}

\section{Introduction}

Current practices in data analytics and artificial
intelligence (AI) perform tasks such as object classification on unending streams of
data.
Computing infrastructure for classification is predominantly oriented toward ``dense''
compute, such as matrix computations.
The continuing exponential growth in generated data\cite{Reinsel2018}
has shifted compute to offload to graphics processors (GPUs) and other focused accelerators
across multiple domains that are dense-compute dominated.

\begin{figure}[tb]
	\centering
    \centering
    \begin{tabular}{m{0.1\columnwidth} m{0.9\columnwidth}}
    (a) & \includegraphics[width=0.55\columnwidth]{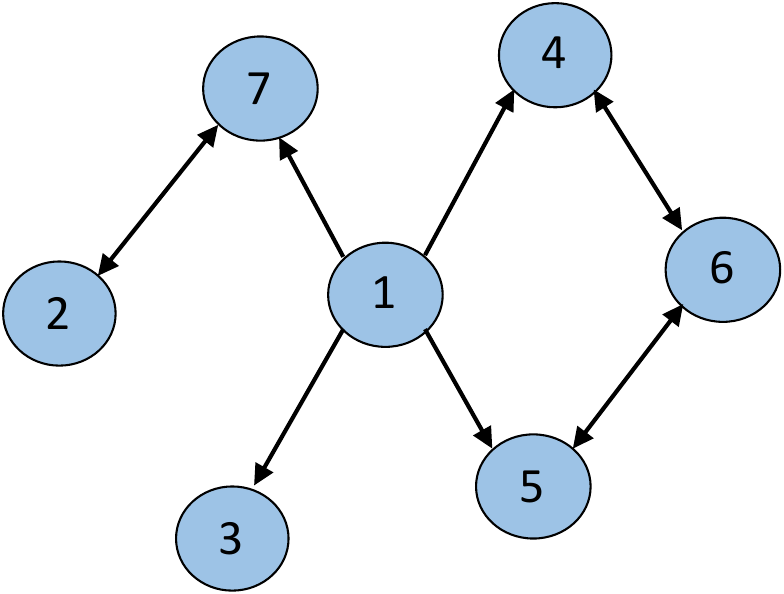}\\
    (b) & \includegraphics[width=0.6\columnwidth]{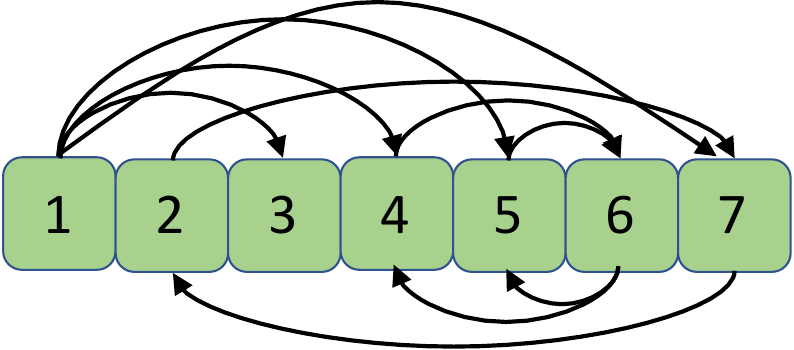}\\
    \end{tabular}
  \caption{
    (a) A sparse graph with directed edges,
    (b) Memory access patterns observed when moving data along the edges of (a)
  }
	\label{fig:graph}
\end{figure}

However, the next step in both AI and data analytics is reasoning about the
\emph{relationships} between these classified objects, typically represented as a graph.
Determining the relationships between entities in a graph is the basis of \emph{graph analytics} \cite{graphsbigdata}.
Graph analytics poses important challenges on existing processor architectures due to its sparse structure.
This sparseness leads to scattered and irregular memory accesses and communication, challenging the optimizations implemented for decades that have
gone into traditional dense compute solutions.
Consider the common case of \emph{pushing} data along the graph edges, see the example graph in \cref{fig:graph}.
All vertices initially store a value locally and then proceed to add their value to all neighbors along outgoing edges. 
This basic computation is ubiquitous in graph algorithms such as \emph{PageRank}~\cite{page1999pagerank}.
The resulting access stream (\cref{fig:graph}b) is irregular and has no locality, making conventional prefetching and caching useless.

Traditionally, algorithmic analysis, with its Random Access Machine model and big O notation, is based on the principle that
compute is precious and communications are free. Yet, today's physics of implementation shows mostly the opposite.
This is especially relevant for at-scale problems in AI and high-performance computing (HPC), which are showing this trend clearly,
exhibiting very low utilization on ``classical'' dense architectures.
The combination of low performance and very large graph sizes limits the practical use of graph analytics.
Recognizing both the increasing importance of this field, and the need for vastly improved sparse computation performance
compared to traditional approaches, DARPA launched their Hierarchical Identify Verify Exploit (HIVE) program
 to achieve at least 1000$\times$ Performance/Watt breakthrough on
such large problems before the end of 2022 \cite{DARPAHIVE}. 

This paper introduces Intel's response to this challenge with its design called Intel\textsuperscript{\textregistered}~Programmable Integrated Unified Memory Architecture (PIUMA).
The PIUMA machine is
designed for graph analytics at massive scales. PIUMA enables high-performance
graph processing by addressing limitations across the network, memory, and
compute architectures that typically limit performance on graph workloads.

\section{Challenges}
\label{sec:wl_char}

\begin{figure}[t]
	\centering
	\includegraphics[width=\linewidth]{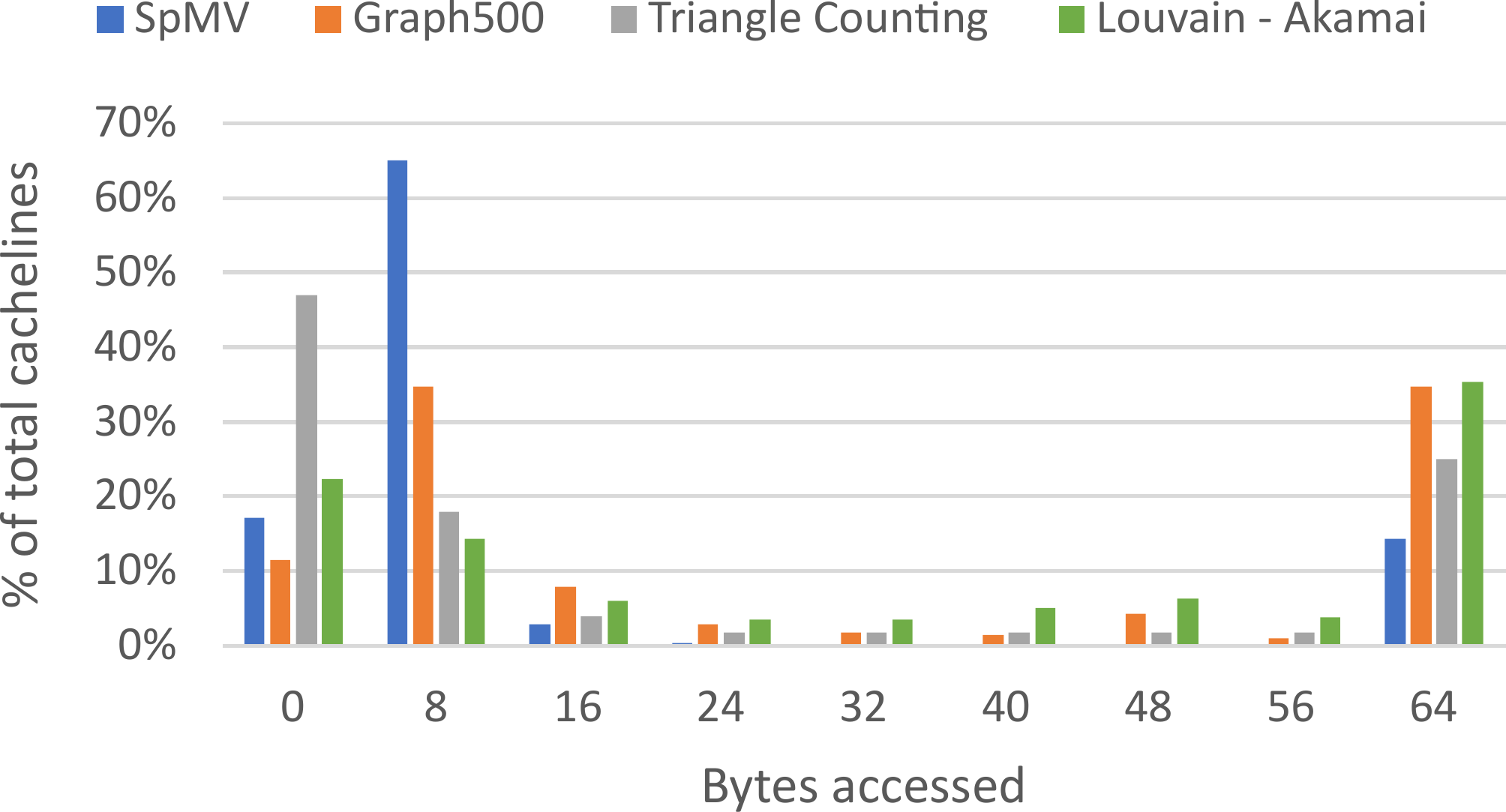}
	\caption{\label{fig:cacheline}Cache line utilization for graph workloads shows a combination of useless prefetches (0B used),
                                  sparse accesses (8B), and streaming data (64B)}
\end{figure}

Graph algorithms face several major scalability challenges on existing architectures, because of their irregularity and sparsity.

\subsection{Challenge 1: Cache and Bandwidth Utilization}
\label{subsec:cl_bw}

Graph analysis applications, when executed on a conventional cache based processor with prefetcher, typically waste a large fraction
of main memory bandwidth\cite{tacosubline}.
For every 64-byte cache line fetched from memory, often just eight bytes or less are used because many data loads are sparse word-sized accesses with no spatial locality.
A typical pattern in graph applications is a chain of indirect loads \cite{LLVMindirects}, similar to a pointer chasing
pattern: a vertex's neighbors are stored in a list, which are used to index the data array.
Since neighbor lists do not show regularity or locality, accesses to the data array are intrinsically sparse.
Other memory access behaviors exhibit increased locality (\eg fetching the neighbor list itself), leading to spatial (but no temporal) cache line reuse.
These lists are limited in size, causing a high rate of useless prefetches that extend past the end of the list.

\cref{fig:cacheline} shows the cache line utilization for a variety of graph analysis applications when executed on a conventional cache based processor with prefetcher.
For every 64-byte cache line fetched from memory, the graph shows how many bytes are actually used by the processor.
For most cache lines, either 0, 8, or the full 64~bytes are used.
The zero usage fraction stems from cache lines that were prefetched but never used.
Cache lines with 8 or fewer bytes used are caused by sparse accesses with no spatial locality.

As a result, the execution of graph applications suffers from inefficient cache and bandwidth utilization: caches
are thrashed with single-use sparse accesses and useless prefetches, and most of the 64 byte memory fetches contain
only one 8-byte useful data element.
Over-provisioning memory bandwidth and/or cache space to cope with sparsity is inefficient in terms of power consumption, chip area and I/O pin count.
Instead, PIUMA uses limited caching and small granularity memory accesses to efficiently deal with the memory behavior of graph applications.

\subsection{Challenge 2: Irregular Computation and Memory Intensity}
Further analysis of graph algorithms shows additional problems in optimizing performance.
The computations are \emph{irregular}: they exhibit skewed compute time distributions, encounter frequent control flow instructions, and perform many memory accesses. 
The compute time for a vertex in the PageRank example is proportional to the number of outgoing edges (degree) of that vertex.
Graphs such as the one illustrated in \cref{fig:graph} have skewed degree distributions, and thus the work per vertex
has a high variance, leading to significant load imbalance.

Analysis reveals that graph applications are heavy on branches and memory operations\cite{scgraphwla}.
Furthermore, conditional branches are often data dependent, \eg checking the degree or certain properties of vertices, leading to irregular and therefore hard to predict branch outcomes.
Together with the high cache miss rates caused by the sparse accesses, conventional performance oriented out-of-order processors are largely underutilized:
data dependencies between cache misses limit the amount of instruction-level parallelism, while hard-to-predict data-dependent branches restrict the amount of useful speculation\cite{eyerman2021enabling}.
In PIUMA, this observation was the incentive to use single issue in-order pipelines with many threads to hide memory latency and avoid speculation.


\subsection{Challenge 3: Fine- and Coarse-Grained synchronization}
Graph algorithms require frequent fine- and coarse-grained synchronization. For example, PageRank requires fine-grained
synchronizations (\eg atomics) to prevent race conditions when pushing values along edges. Synchronization instructions
that resolve in the cache hierarchy place a large stress on the cache coherency mechanisms for multi-socket systems,
and all synchronizations incur long round-trip latencies on multi-node systems. Additionally, the sparse memory accesses
result in even more memory traffic for synchronizations due to false sharing in the cache coherency system.

Coarse-grained synchronizations (\eg system-wide barriers and prefix scans) fence the already-challenging computations
in graph algorithms. These synchronizations have diverse uses including resource coordination, dynamic load balancing,
and the aggregation of partial results. These synchronizations
can dominate execution time on large-scale systems due to high network latencies and imbalanced computation.

\subsection{Challenge 4: Massive Datasets}

Current commercial graph databases exceed 20 TB as an in-memory representation.
Such large problems exceed the capabilities of even a rack of computational nodes of any type and would require a
large-scale multi-node platform to even house the graph's working set.
When combined with the prior observations---poor memory hierarchy utilization, high control flow changes,
frequent memory references, and abundant synchronizations---any architecture that targets graph workloads
must focus on reducing latency to access remote data, combined with latency hiding techniques in the processing elements.

\begin{figure*}[t]
	\centering
	\includegraphics[width=.9\textwidth]{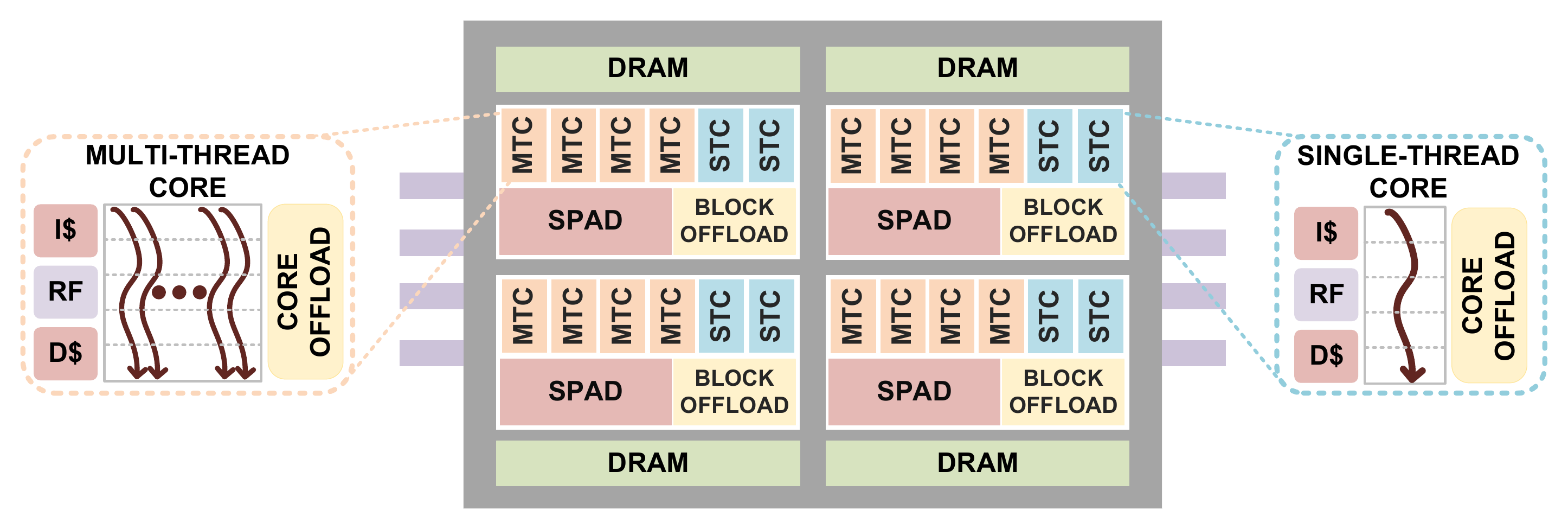}
	\caption{High-level diagram of the PIUMA architecture}
	\label{fig:puma_arch_overview}
\end{figure*}

\vspace{3mm}

Although the analysis in this section focuses on CPUs, the same challenges apply for GPUs: sparse accesses prevent memory
coalescing, branches cause thread divergence and synchronization limits thread progress.
Nevertheless, for small graphs, GPUs usually perform better on graph algorithms than CPUs for small graphs~\cite{simd-x} because they have
more threads, which hides memory latency, and much higher memory bandwidth, brute-forcing the inefficient bandwidth utilization.
However, GPUs also have limited memory capacity and scale-out capabilities, which means that they are unable to process large, multi-TB graphs.
Furthermore, graphs are extremely sparse ($\ll$1\% non-zeros) \cite{RMAT}, so the typical GPU trick to densify the adjacency
matrix for an efficient GPU execution leads to another few orders of magnitude increase in memory usage, restricting it to small graphs only\cite{burtscher2012quantitative}.
PIUMA directly operates on sparse data (\eg compressed sparse row (CSR) format) which avoids the need for densification.

\section{Introducing PIUMA}


The observations on graph analysis workloads guided the PIUMA design, targeting breakthrough performance per Watt for graph analytics.
We discuss how each component of the PIUMA architecture is designed to cope with the challenges imposed by graph workloads.

\subsection{PIUMA Cores}

The design of PIUMA cores builds on the observations that most graph workloads have abundant parallelism, are memory bound, and are not compute intensive.
These observations call for many simple pipelines, with multi-threading to hide memory latency.~\cref{fig:puma_arch_overview} shows a block diagram of the PIUMA archirecture.
PIUMA multi-threaded cores (MTC) are round-robin multi-threaded in-order pipelines~\cite{barrel}.
At any moment, each thread can only have one in-flight instruction (implicitly making them stall-on-miss at the thread level), which considerably simplifies the core design for better energy efficiency.
Single-threaded cores (STC) are used for single-thread performance sensitive tasks, such as memory and thread management (\eg from the operating system).
These are in-order stall-on-use cores that are able to exploit some instruction and memory-level parallelism, while avoiding the high power consumption of aggressive out-of-order pipelines.
Both core types implement the same custom RISC instruction set which enables easy thread migration.

Each core has a small data and instruction cache (D\$ and I\$), and a large register file (RF) with 32~registers per thread.
Because of the low locality in graph workloads, no higher cache levels are included, avoiding useless chip area and power consumption of large caches.
Special cache instructions are built into the ISA to support I\$ and D\$
software prefetches, invalidations, and write-backs.
A MOESI-F protocol \cite{piuma_moesif} maintains coherency across all data-caches on the die,
and allows for migration of dirty cache lines between data caches.
For scalability, caches are not coherent across the whole system.
It is the responsibility of the programmer to avoid modifying shared data that are cached, or to flush caches if required for correctness.

MTCs and STCs are grouped into \emph{blocks}, each of which has a large local scratchpad (SPAD) for low latency storage.
Programmers can manipulate address bits to point to specific memory map regions (scratchpad, main memory, configuration registers, etc.).
A cache bit determines whether the memory access is cached. The runtime language can choose to \eg cache the execution stack and not cache static data by default.
There are no hardware prefetchers to avoid useless data fetches and to limit power consumption.
Instead, the offload engines described below can be used to efficiently fetch large chunks of useful data.

\subsection{Offload Engines}

Although the MTCs hide much of the memory latency by supporting multiple concurrent threads, their in-order design limits the number of outstanding memory accesses to one per thread.
To increase memory-level parallelism and to free more compute cycles to the cores, multiple memory offload engines are added to each block.
The offload engine performs memory operations typically found in many graph applications in the background, while the cores continue with their computations.

The direct memory access (\emph{DMA}) engine performs operations such as (strided) copy, scatter/gather.
This engine has the capability to interpret various compressed sparse representations commonly used in neighbor lists (\eg CSR) and perform other data transformations
(\eg transpose, basic arithmetic operations, etc.).

\emph{Indirect operations} accelerate pointer chasing (\eg {\tt A[B[i]]}). Conventionally, this is done in three steps:
an offset {\tt B[i]} is loaded from memory into a register, the load/store address is computed by the core by shifting the loaded offset and adding it to the base address {\tt A},
and finally the load or store operation is performed on the created address.
ISA support for indirect loads and stores\cite{LLVMindirects} eliminates the need to bring the offset to the core by
performing these three steps on the fly.
We add a special network message type to support indirect loads, containing an index address {\tt \&B[i]},
base address value {\tt A} and the operation to compute the final load target (\eg {\tt A+8*B[i]} for a 64-bit index).
When the index and data values are located on the same memory controller, the round-trip that would normally send {\tt B[i]}
back to the core to calculate {\tt \&A[B[i]]} is completely avoided. When the {\tt A} and {\tt B} arrays reside at different memory controllers,
the intermediate address is sent directly from one memory controller to the next one, so we require three instead of four network traversals.

\emph{Queue} engines are responsible for maintaining queues allocated in shared memory, alleviating the core from atomic inserts and removals.
They can be used for work stealing algorithms and dynamically partitioning the workload.
\emph{Collective} engines implement efficient system-wide reductions and barriers.
\emph{Remote atomics} perform atomic operations at the memory controller or scratchpad where the data is located, instead of burdening the pipeline with
first locking the data, moving the data to the core, updating it, writing back and unlocking.
They enable efficient and scalable synchronization, which is indispensable for the high thread count in PIUMA.

The engines are directed by the PIUMA cores using specific instructions.
These instructions are non-blocking, enabling the cores to perform other work while the operation is done in the background.
Custom polling and waiting instructions are used to synchronize the threads and offloaded computations.

\subsection{Memory Organization}

Sparse and irregular accesses to a large data structure are typical for
graph analysis applications.
Therefore, accesses to remote memory should be done with minimal overhead.
PIUMA implements a hardware distributed global address space (DGAS), which enables each core to uniformly access memory across the full
system with one address range.
Besides avoiding the overhead of setting up communication for remote accesses, a DGAS also greatly simplifies programming, because
there is no implementation difference between accessing local and remote memory.
Address translation tables (ATT) contain programmable rules to translate application memory addresses to physical locations, to arrange 
the address space to the need of the application (\eg address interleaved, block partitioned, etc.).

The memory controllers (one per block) are redesigned to support native 8-byte accesses, while supporting standard cache line accesses as well.
Fetching only the data that is actually needed reduces memory bandwidth pressure and utilizes the available bandwidth more efficiently.

\subsection{Network}

The network connecting the blocks is responsible for sending memory requests to remote memory controllers.
Similar to the memory controller, it is optimized for small 8-byte messages.
Furthermore, due to the high fraction of remote accesses, network bandwidth exceeds local DRAM bandwidth,
which is different from conventional architectures that assume higher local traffic than remote traffic.

To obtain high bandwidth and low latency to remote blocks, the network needs to have a high radix and a low diameter.
This is achieved with a HyperX topology~\cite{hyperx}, a hierarchical network with all-to-all connections on each level.
To ensure power-efficient, high-bandwidth communication, optical links are used that are tightly integrated into the PIUMA chip package.
The hierarchical topology and optical links enable PIUMA to efficiently scale out to many nodes, maintaining easy and fast remote access.

\subsection{Comparison to other Graph Processors}

The Cray Urika-GD graph processor~\cite{threadstorm} was one of the first commercial graph-oriented big data processors.
Similar to PIUMA, it consisted of multiple many-threaded cores with no large caches and a memory-coherent network.
It did not support fine-grained 8-byte accesses, wasting bandwidth on loading full cache lines.
Furthermore, it had no offload memory engines, such as the DMA, queue and remote atomics in PIUMA, leading to more memory stalls in the pipelines.

The Emu architecture~\cite{emu} is a recently proposed architecture for big data analysis, including graph analysis workloads.
Similar to PIUMA and Urika-GD, it consists of many small cores with many hardware threads per core to hide memory latency.
It also features 8-byte DRAM accesses and is completely cacheless.
Unique is its low-overhead thread migration scheme, which enables moving threads to a core near to the memory controller that owns the required data instead of moving the data to the current core.
Moving threads to data is beneficial if the overhead of moving the thread is compensated by the amount of locally consumed data.
Young \etal~\cite{young} report that migrating a thread involves moving 200~bytes, which means that at least 25 local 8-byte accesses are needed to compensate for the thread migration.
Therefore, optimizing data locality is crucial for obtaining good performance on Emu~\cite{young}, which is often hard to obtain for graph analysis applications.
In contrast, PIUMA does not rely on any locality.
Instead, it uses the offload engines to perform complex system-wide memory operations in parallel, and only moves the data that is eventually needed to the core that requests it.
For example, a DMA gather will not move the memory stored indices or addresses of the data elements to gather to the requesting core, only the requested elements from the data array are moved.

Song \etal~\cite{song} propose a graph processor based on sparse matrix algebra, building on the observation that many graph applications can be represented as operations on sparse matrices.
Their architecture has overlaps with PIUMA, such as the absence of caches, and fine-grained communication and memory accesses.
Graphicionado~\cite{graphicionado} is a graph analysis accelerator, implementing a vertex-centric compute paradigm.
While these accelerators are likely more energy efficient for analyzing small graphs, PIUMA's goal is to provide a flexible instruction set architecture,
optimized for typical graph analysis operations, and not to be limited to algorithms that use sparse matrix algebra or vertex-centric operations.
Furthermore, none of these proposals scale out to multi-TB graphs with trillions of vertices.

\section{Software stack}
\label{sec:software}



The PIUMA software stack includes all the tools necessary for
developers to write, compile, execute, and debug codes.
Unlike most accelerators, the PIUMA hardware and its programming stack
do not impose upon the programmer a restricted parallel programming model, derived from
hardware limitations. Instead, the software stack
offers flexibility in leveraging host and accelerator resources,
and takes a layered programming abstraction approach. Each abstraction layer offers
a trade-off between programmer control and productivity, catering to
a range of developer audiences.

Standalone x86 C/C++ workloads are usually straightforward to port to
the PIUMA software stack, often simplifying the code in the process by
removing the need for complex locality-improving datastructures.

The following sections highlight the PIUMA software development
kit and three programming interfaces to target PIUMA:
a C-based single program, multiple data (SPMD) programming model, a PIUMA implementation of OpenMP,
and plugins for Anaconda's Metagraph framework in Python.

\subsection{PIUMA Software Development Kit}

\begin{figure}[t]
  \centering
  \includegraphics[width=.8\linewidth]{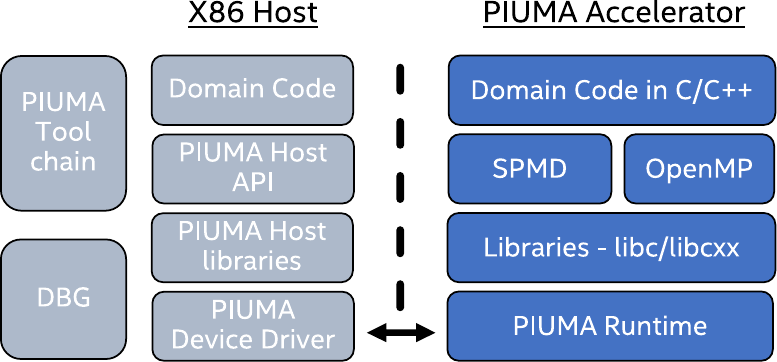}
  \caption{High-level diagram of the PIUMA software stack}
  \label{fig:piuma_sw_stack}
\end{figure}

The PIUMA Software Development Kit (SDK) provides developers with
a fully featured software stack composed of a set of familiar C and C++ programming
interfaces, toolchain, runtime, driver, supporting libraries,
and debugger.
\cref{fig:piuma_sw_stack} shows the general organization of
the software stack, spanning host and accelerator.

The SDK tools and libraries make extensive use of PIUMA
hardware-backed features such as atomics, queues, collectives, and DMAs.
Additionally, some of the hardware features are automatically leveraged by
the toolchain. For example, code can be compiled to automatically make use
of PIUMA's indirect-load~\cite{LLVMindirects} and bitwise instructions.
Developers can also directly use PIUMA-specific compiler builtins
to access custom RISC ISA instructions.

Since the PIUMA accelerator is deployed alongside a host, developers
decide whether to program PIUMA in standalone or hybrid mode.
In standalone mode, the full program is compiled to run natively on
PIUMA. The LLVM-based PIUMA toolchain contains the expected suite of
tools: a compiler, binary utilities (assembler, linker, ELF tools,
etc.), and support libraries (libc, libcxx, libunwind, etc.) tailored
for PIUMA.

PIUMA-native binaries can be launched as POSIX-style processes through
a convenience launcher tool on the x86 Host. This launcher
relies on more general user-space libraries (Host API) providing
management, communication, and debugging support on top of the PIUMA
device driver. On the accelerator's side, the program is handed off to
the PIUMA runtime, which manages hardware resources and program
executions, fulfilling the role of a light-weight operating system.

In hybrid mode, the main program runs on x86 and coordinates the
execution of tasks or kernels with the PIUMA accelerator using the
Host-API. The Host-API covers basic functionalities such as dynamically
allocating memory, executing data transfers in and out, as well as
scheduling parallel functions for execution.

\subsection{Programming Interfaces}

Most programmers will want to use one of the supported parallel
programming interfaces: a single program, multiple data (SPMD) programming model, OpenMP, or graph
algorithms plugins for Anaconda's Metagraph.

The PIUMA SPMD programming model bridges the gap between system-level
programming and application-level programming, implementing enough of
a runtime and OS-like functionality to offer the user a familiar
(shared-memory) system view and parallel programming approach
(bulk-synchronous style). The PIUMA SPMD layer is a good target for
developers that require low overheads and tight control, as it allows
the underlying runtime to scale well to a large number of threads. A
PIUMA SPMD program is a standard C program: the {\tt main()} function
is the entry point and is executed sequentially. There, developers can
use the provided SPMD API to execute user-defined functions across the
available hardware threads.  The SPMD programming layer API also
provides a thin library for: thread identification and system geometry
information to orchestrate parallel computations, scratchpad and
global memory allocations, point-to-point synchronization through the
use of atomics and hardware queues, and global synchronization with
hardware collectives.



The OpenMP layer provides developers
with increased productivity and a familiar programming environment.
Some of PIUMA's hardware features
find a natural fit in standard OpenMP pragmas (atomics, reductions)
while others require the creation of new pragmas. For example, a
pragma can be applied to a specific section of code to instruct the
PIUMA compiler to accelerate code with indirect access instructions.

For domain scientists mostly interested in using off-the-shelf graph algorithms,
we have developed PIUMA plugins for Anaconda's Metagraph Python library. This
approach enables domain scientists to leverage PIUMA hardware from a familiar
Python programming environment such as Jupyter notebooks.
When invoking graph algorithms in the Metagraph framework, developers
can use annotations to request the use of a PIUMA implementation.
The Metagraph dispatcher then finds and executes the corresponding PIUMA
plugin, which takes care of all the necessary steps to
transparently offload data and compute from the host to the accelerator.

\section{Simulation infrastructure}
\label{sec:codesign}

Crucial for the pathfinding and development of PIUMA was the hardware/software co-design process.
This process requires the involvement of multiple multi-disciplinary teams: architects, system
software developers, workload analysis teams, performance simulation and analysis teams, as well as FPGA emulation teams.
In parallel with the hardware design and compiler development, we developed an architectural
simulator for PIUMA, simulating the timing of all instructions in the pipelines, engines, memory
and network, based on the hardware specifications.

The functional part of the simulator was created by extending FSim\cite{runnemede} to handle the custom RISC ISA
as well as the functional emulation of the offload engines and network collectives.
For performance modeling, Sniper\cite{carlson2014aeohmcm} was used to model the PIUMA cores, memory subsystem and interconnection network.
In addition to performance estimations of running a workload on PIUMA, it provides an extensive set of performance analysis reports,
such as Cycle per instruction (CPI) stacks and detailed performance information on each memory structure and each instruction.
This enables workload owners to quickly detect bottleneck causes, and to use these insights to
optimize the workload for PIUMA and report hardware bottlenecks to the hardware design team.
The hardware team then responds with an updated design, feeding a continuous cycle of gradual
improvements to hardware and software.
We validated the resulting simulator against the PIUMA FPGA emulation platform as well as against A0 silicon
once that became available.


\subsection{Functional Simulator (FSim)}
FSim is a fast functional simulator developed by Intel\cite{runnemede} for the Open Community Runtime project\cite{OCR}. 
As part of the PIUMA co-design, we massively overhauled the FSim framework to enable quick software
development, analysis and RTL validation for data center scale systems. FSim is a distributed full-system functional simulation framework capable of simulating
thousands of cores using a large cluster of machines. FSim can simulate each core in the single to double-digit millions of instructions per second (MIPS) range of performance (depending on the workload's
compute-to-communicate ratio) given a large enough compute cluster. It can operate in standalone mode, where it runs pre-loaded compiled binaries for the supported ISA
based on an LLVM toolchain. It can also operate in an offload mode, where it can interface with other simulators such as QEMU\cite{qemu} to model a host-device
system (where QEMU emulates an x86 host and FSim simulates PIUMA), Synopsys VCS\cite{vcs} for performing RTL design verification, etc. FSim captures an extensive set of
spatial statistics (instruction histogram, event counters, communication patterns, etc.) in support of workload characterization.

\subsection{Performance Simulator (Sniper)}
Sniper is a next generation parallel, high-speed and accurate x86 simulator which can be easily
extended to simulate new architectures\cite{carlson2014aeohmcm}. This multi-core simulator uses the Pin dynamic instrumentation tool\cite{PIN}
as a functional front-end and has performance models for several core types and cache hierarchies.
Sniper allows for fast and accurate simulation of large (hundreds of cores) multiprocessors, and for trading off simulation speed for accuracy to
allow a range of flexible simulation options when exploring different homogeneous and heterogeneous multi-core architectures.
Because Sniper is a functional-first simulator (with timing feedback to enable accurate multi-core simulation),
it is relatively easy to retarget to non-x86 instruction-set architectures. In fact, a RISC-V based version has recently been made
available publicly\cite{mallya2018flexible}.

We built a timing simulator model for the PIUMA architecture, using FSim as the functional frontend to handle the custom RISC ISA
as well as the functional emulation of the offload engines and network collectives.
The in-order mode of Sniper's instruction-window centric core model was extended to model both the stall-on-use single-threaded (STC) cores
as well as the stall-on-miss multi-threaded (MTC) cores. We also extended the memory model to handle scratchpads, selective caching, and
the HyperX topology of the network.
We validated the resulting simulator against the PIUMA FPGA emulation platform as well as against A0 silicon
once that became available.

\begin{figure}[t]
	\centering
	\includegraphics[width=\linewidth]{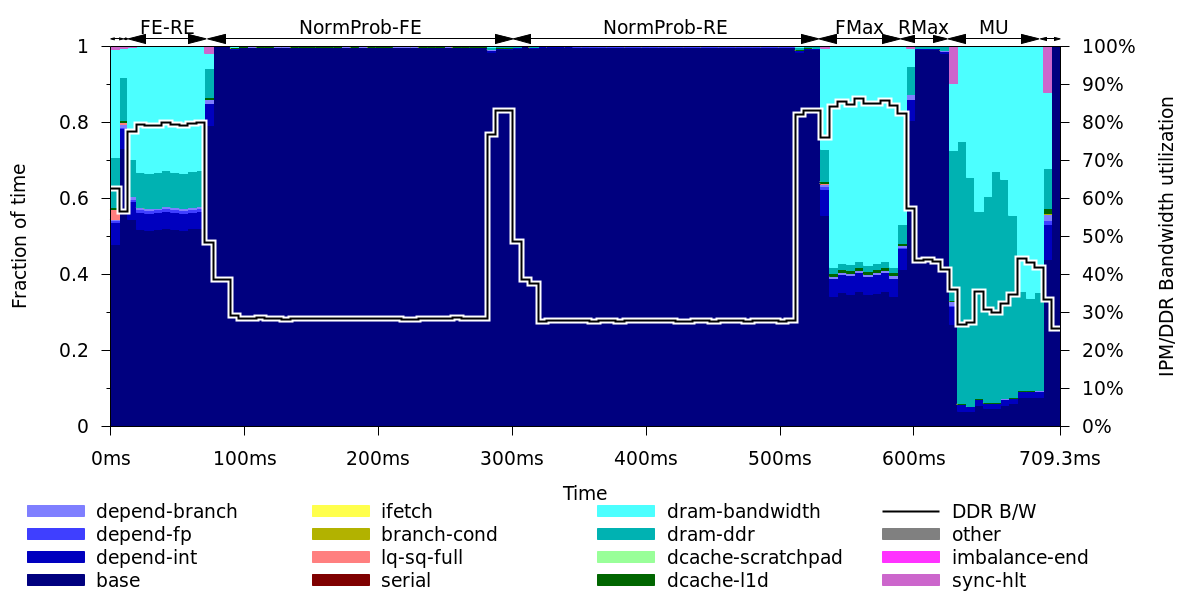}
	\caption{Simulated execution profile of Application Classification for RMAT-18 on 8~PIUMA blocks}
	\label{fig:application_classification_cpi}
\end{figure}

By building on an established timing simulator, all of its performance analysis tools were easily made available for
PIUMA as well. In particular, CPI stacks\cite{heirman2011ucstusbimw} and instruction-level statistics\cite{heirman2015sniper}
were essential for application, compiler and library developers to help them understand an unfamiliar architecture with often surprising performance effects,
and for the architects to gain insight into, and solve, micro-architectural bottlenecks.
\cref{fig:application_classification_cpi} shows an example CPI stack of the Application Classification workload,
which clearly illustrates the different application phases and how some are compute versus memory bound.
Together, these tools allowed the PIUMA team to do extensive co-design of the full hardware and software stack.

\subsection{Multi-Node Simulation}
In addition to simulating a single PIUMA chip (eight blocks running a total of 528~threads), we wanted to validate our analytical multi-chip scaling models
against simulation results as much as possible. Because our PIUMA simulator is built from loosely coupled functional and timing components,
it was possible to run these on different host machines to increase the maximum size of the machine we could model.
Moreover, FSim was already built as an MPI program so it is able to be run in a distributed mode natively.

The functional simulator has to deal with the massive thread count and memory requirements of a workload, therefore, this part runs
on multiple hosts using MPI communication. Simulated compute and memory are distributed across multiple hosts to maximize simulator performance.
Each execution thread in FSim sends a dynamic instruction stream over an inter-process communication channel (TCP socket) to the timing
model (Sniper). The timing model itself has lower memory requirements since it only needs to store timing-relevant microachitectural state
(which in case of PIUMA is relatively limited because of small caches, absence of speculation, etc.), yet requires fine-grained synchronization between different
threads (cores and memory communicate with each other at nanosecond time scales). Therefore, the timing model runs as one
multi-threaded application on a single host to avoid the large synchronization bottleneck that would be incurred if the timing simulation was to be spread across multiple machines.
With this setup, we were able to run simulations of architectures up to 256 PIUMA blocks (32 chips running 16,896 threads).

\subsection{FPGA Emulation}

We also made extensive use of FPGA emulation to verify the RTL design. The PIUMA FPGA platform has been deployed in parallel to workloads, simulators and hardware
development efforts for both functional validation and performance correlation. The FPGA model was always generated from the latest RTL repository, contained
all hardware hierarchies and included third party IP. For correctness verification, we ran multiple tests from tiny codes to
large workloads to make sure the FPGA bitstreams are working and are in sync with the concurrent developments of the RTL and simulators.
This allowed the RTL teams to fix over 50 critical bugs before tape-out.
With this co-design approach we evaluated performance progression and regressions between software and hardware.

\section{PIUMA prototype implementation}
\label{ssec:results:hardware}

\begin{figure}[t]
	\centering
	\includegraphics[width=\linewidth]{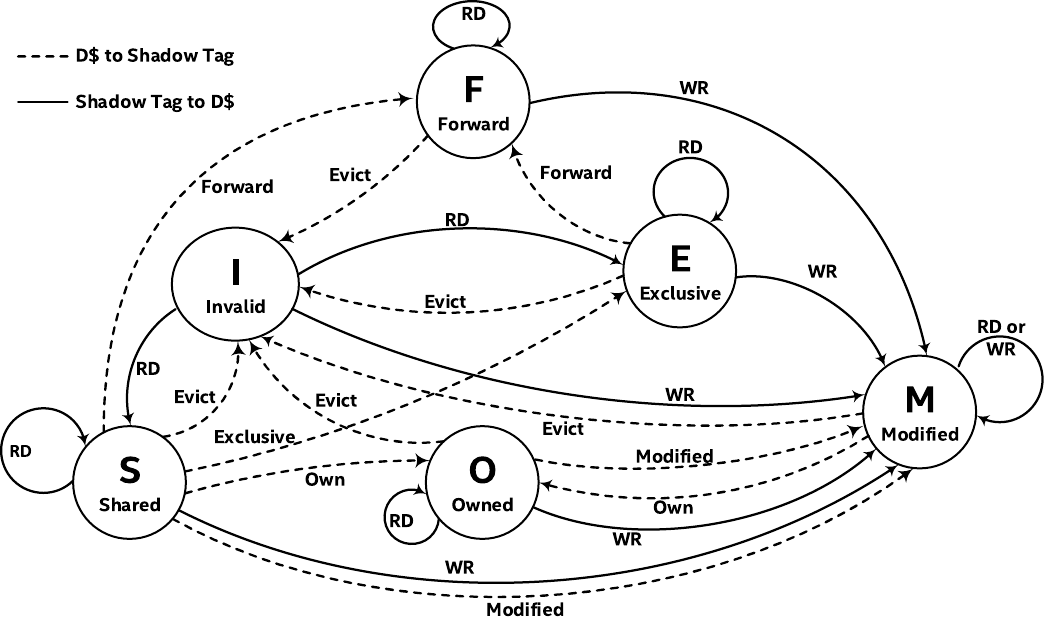}
	\caption{MOESI-F coherency protocol state diagram}
	\label{fig:moesif}
\end{figure}

We designed and fabricated an A0 test chip
that implements the PIUMA architecture. This chip was powered on and characterized in the lab,
and was healthy enough (with some software mitigations in place) to run actual workloads.
We also demonstrated optical connectivity between a pair of PIUMA chips.
Finally, a 16-chip PIUMA {\em node} was constructed to allow validation of the inter-chip
network and conduct scaling experiments.

\subsection{A0 hardware design}

We designed and built a 27.6B-transistor, 316mm$^2$ prototype chip in 7nm FinFET CMOS~\cite{tsmc7nm}.
The chip integrates eight PIUMA blocks (running a total of 528 threads),
32MB of on-die scratchpad memory, and all off-chip interfaces (memory, network, and a PCIe link to the host system).

A MOESI-F protocol \cite{piuma_moesif} maintains coherency across all data caches on the die.
The state diagram (see \cref{fig:moesif}) allows for migration of dirty cache lines between data caches
and ensures that the data will not be evicted from the coherent domain until no sharers exist.
A die-level shadow tag is used to track the current state of all cache lines.

Memory transactions are distributed over eight narrow-channel on-die DDR5 controllers
that have been optimized for 8-byte native accesses for efficient bandwidth
utilization on sparse graph data sets.
Main memory is built from standard DDR5 memory chips in a custom SODIMM form factor
that connects each memory chip directly to its own memory controller using a dedicated command/address bus.
Each of the eight on-die memory controllers can be accessed in parallel at 
an aggregate peak memory bandwidth of
35.2GB/s. 

\begin{figure}[t]
	\centering
	\includegraphics[width=\linewidth]{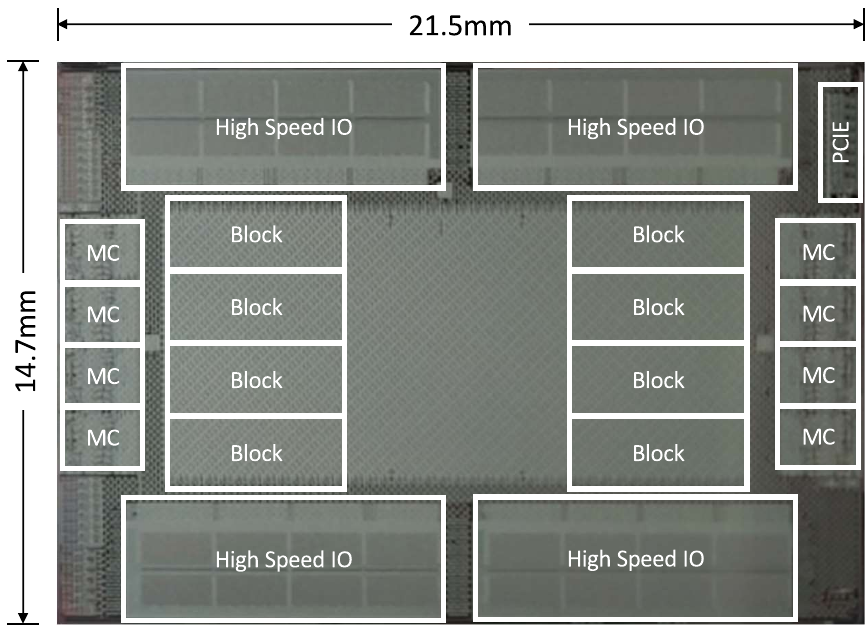}
	~\\
	\footnotesize
	\begin{tabular}{@{ }l@{ }l@{ }}
		\toprule
		Technology & TSMC 7nm FinFET \\
		Interconnect & 15 metal layers \\
		Die transistors & 27.6B \\
		Die area & 316~mm$^2$ \\
		Block transistors & 1.2B \\
		Block area & 9.3~mm$^2$ \\
		Signals & 705 \\
		Package & 3275-pin BGA \\
		\bottomrule
	\end{tabular}
	\caption{Full-chip micrograph and characteristics}
	\label{fig:isscc_dieshot}
\end{figure}

A 10-port virtual cut-through router 
used to create the
on-die network employs a credit-based flow control protocol.
Router ports are packet-switched, have 25-byte data links, and can
operate at 1~GHz. Optimized for graph analysis, packets sizes are
1, 2, or 4 flow control units resulting in a maximum data payload
sizes of 8-byte, 16-byte, or 64-byte, respectively.
No-load router latency is 4 clock cycles, including link traversal.
The on-die network uses a 2-dimensional mesh to connect routers.
An XY dimension ordered routing algorithm is loosely followed and
individual router links offer 64~GB/s interconnect bandwidth,
enabling the on-die network to support 1~TB/s of bisectional bandwidth.

Four x8 high speed I/O links are divided into 32~channels for 1~TB/s per direction
of off-die signaling.
A PCIe Gen 4 x8 controller within each
die delivers 16GB/s of additional bandwidth and
provides an endpoint device to a conventional x86 host system.

\cref{fig:isscc_dieshot} shows the layout of the chip.
The most critical resource was the edges of the chip (shoreline) required to fit the memory controllers
and off-chip network interfaces. Much of the internal area is taken up by routing of the on- and off-die networks
while the cores and memories take up just 24\%.


\begin{table}[t]
	\caption{PIUMA system hierarchy}
	\label{tab:system_hierarchy}
	\centering\footnotesize
	\begin{tabular}{lp{.8\linewidth}}
		\toprule
		{\em Block} & 66 hardware compute threads; 192KB instruction+data cache; 4MB scratchpad SRAM; uncore engines \\
		{\em Socket} & 8 blocks; 32 optical I/O ports at 32 GB/s/dir each; 32GB custom DDR5-4400 DRAM; PCIe G4 x8 \\
		{\em Node} & 16 sockets in an Open Compute Project (OCP) sled form factor, 0.5 TB DRAM, 16 TB/s/dir optical bandwidth \\
		\bottomrule
	\end{tabular}
\end{table}

\begin{table}[t]
	\caption{Configurations and bandwidths across PIUMA system sizes for HyperX topologies}
	\label{tab:configurations}
	\centering\resizebox{\columnwidth}{!}{%
	\begin{tabular}{c|c|ccc|ccc|c}
		\toprule
		\multirow{2}{*}{Nodes} & \multirow{2}{*}{HyperX Levels} & \multicolumn{3}{c|}{Configuration} & \multicolumn{3}{c|}{Ports/node pair} & \multirowcell{2}{Uni-directional \\ bisection b/w (TB/s)} \\
		& & $n_0$ & $n_1$ & $n_2$ & $m_0$ & $m_1$ & $m_2$ & \\
		\midrule
		2	& 1	& 2	 & & &	 	128	 	& & & 	1 \\
		4	& 1	& 4	 & & &	 	64	 	& & & 	2 \\
		8	& 1	& 8	 & & &	 	32	 	& & & 	4 \\
		16	& 1	& 16 & & &		 	16	 & & &	 	8 \\
		32	& 1	& 32 & & &	 	 	8	 & & &	 	16 \\
		64	& 1	& 64	 & & &	 	4	 & & &	 	32 \\
		128	& 1	& 128	& & & 	 	2	 & & &	 	64 \\
		256	& 1	& 256	& & & 	 	1	 & & &	 	128 \\
		512	& 2	& 32	& 16	 & & 	4	& 8	 	& & 128 \\
		1,024	& 2	& 32	& 32	 	& & 4	& 4	 	& & 256 \\
		2,048	& 2	& 64	& 32	 	& & 2	& 4	 	& & 512 \\
		4,096	& 2	& 64	& 64	 	& & 2	& 2	 	& & 1,024 \\
		8,192	& 2	& 128	& 64	 	& & 1	& 2	 	& & 2,048 \\
		16,384	& 2	& 128	& 128	 	& & 1	& 1	 	& & 4,096 \\
		32,768	& 3	& 32	& 32	& 32	& 2	& 2	& 2	& 4,096 \\
		65,536	& 3	& 64	& 32	& 32	& 2	& 2	& 2	& 8,192 \\
		131,072	& 3	& 64	& 64	& 32	& 2	& 1	& 2	& 16,384 \\
		\bottomrule
	\end{tabular}
	}
\end{table}

\begin{figure}[t]
	\centering
	\includegraphics[width=\linewidth]{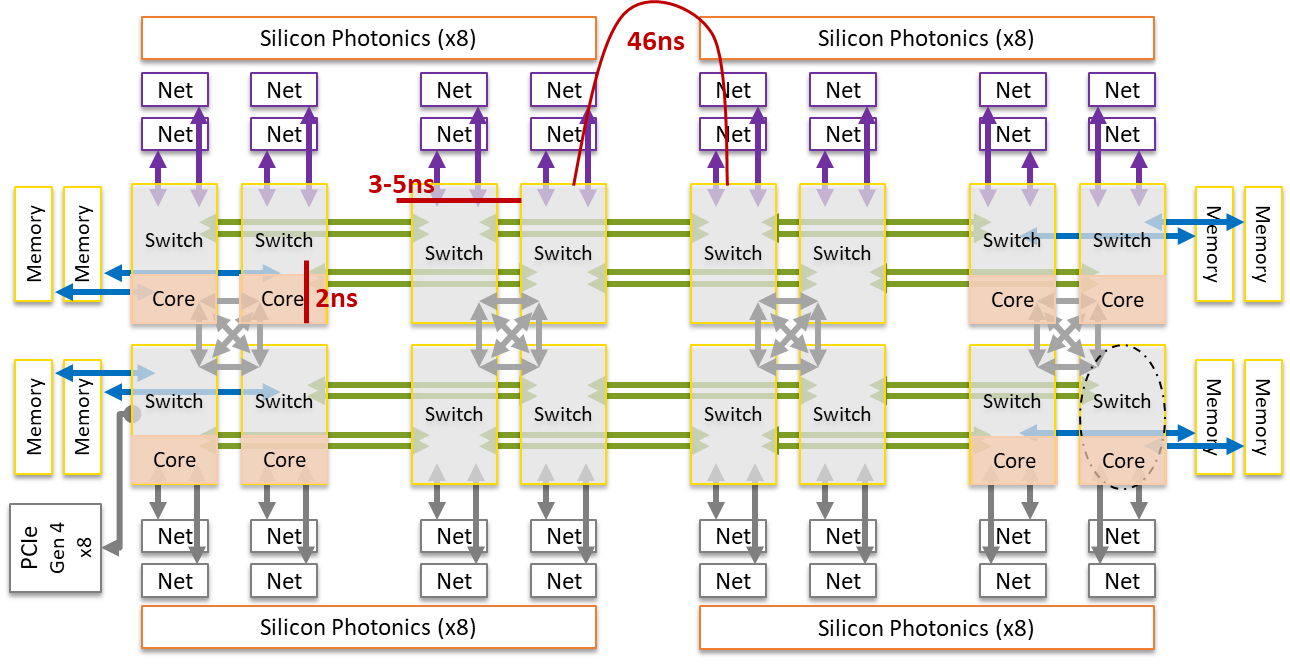}
	\caption{Block diagram of a single PIUMA socket with latency costs}
	\label{fig:hotchips_latencies}
\end{figure}

\subsection{Inter-chip network}

To address the need to scale from small to extremely large (1-10 PB DRAM) machines,
while being hypersensitive to latency in scaling performance, PIUMA fused traditional network switch logic
into the compute logic to make a disaggregated switch fabric in a HyperX topology\cite{hyperx} that makes
a glueless connection system between all PIUMA sockets.
Utilizing the latest co-packaged optical (CPO)  technology, PIUMA converts the on-die mesh protocol directly to the optical fabric
and back to seamlessly glue together all PIUMA chips in a system in a large virtual die\cite{hotpiuma}.
All routing is done by the global address associated to each resource in the system---no protocol such as Ethernet or CXL is required.

All PIUMA socket-to-socket optical I/O (OIO) links utilize co-packaged optical (CPO) silicon photonics chiplets
to overcome the limitations of electrical signaling, board design, and routing complexity.
These chiplets connect via the Advanced Interface Bus (AIB) 1.0 protocol over Intel's
embedded multi-die interconnect bridge (EMIB).
A custom IP block encapsulates the PIUMA mesh protocol over EMIB, AIB, and OIO interfaces.

The PIUMA system fabric topology is differentiated by the connectivity of the sockets.
A {\em node} is a group of 16 sockets with all-to-all OIO connectivity,
providing 512GB of DRAM, 16 TB/s/dir over 512~OIO ports, and $>$8k hardware threads, see \cref{tab:system_hierarchy}.
In this specific configuration, 15~OIO ports per socket support intra-node connections with
a single OIO link (32 GB/s/dir) traversal at $<$50ns creating direct compute die mesh-to-mesh traffic
without intervening protocols or translations. As the system scales to the node level and beyond,
the remaining 17~OIO links per PIUMA socket enable a HyperX topology (see \cref{tab:configurations}).
This connectivity approach allows PIUMA to scale to two million sockets per system
and maintains an increasing rate of interconnect bisection bandwidth accordingly.

To provide an example of the worst-case one-way latency within a node, assuming 16 sockets all-to-all connected,
the worst-case latency is determined by the longest on-die route.
For this example, the source socket routing block~0 to block~7's OIO port,
and the destination socket routing block~0's OIO port to block~7 is the worst-case latency,
as shown in \cref{fig:hotchips_latencies} by using the maximum (five) mesh hops per socket.
A majority of the worst-case latency is spent on the socket's mesh network at both the source and destination:
68ns in on-die routing. The encoding latency for AIB is $\sim$72\% of the OIO time
in the current unoptimized first prototype implementation.
Early analysis for optimized timings indicates that one-way latencies can be best-case $<$17ns and worst-case $<$45ns.

\subsection{A0 power-on}
PIUMA A0 power-on was done in four phases.
The first phase was to validate the platform with the PIUMA A0 chip installed.
The second phase was to bring the chip out of reset and verify clocking using built in monitor circuitry.
The third phase was focused on validating the functionality of the different IP blocks.
The fourth phase focused on running functional test contents for the PIUMA chip.
It took less than six weeks to start a PIUMA functional part from the arrival of the first silicon.

We have a healthy PIUMA A0 chip compared to other chips of this scale on a brand new architecture
and a grounds up new design. Few hardware bugs have been uncovered,
although some required software workarounds with significant performance impact.

\begin{figure}[t]
	\centering
	\includegraphics[width=.49\linewidth]{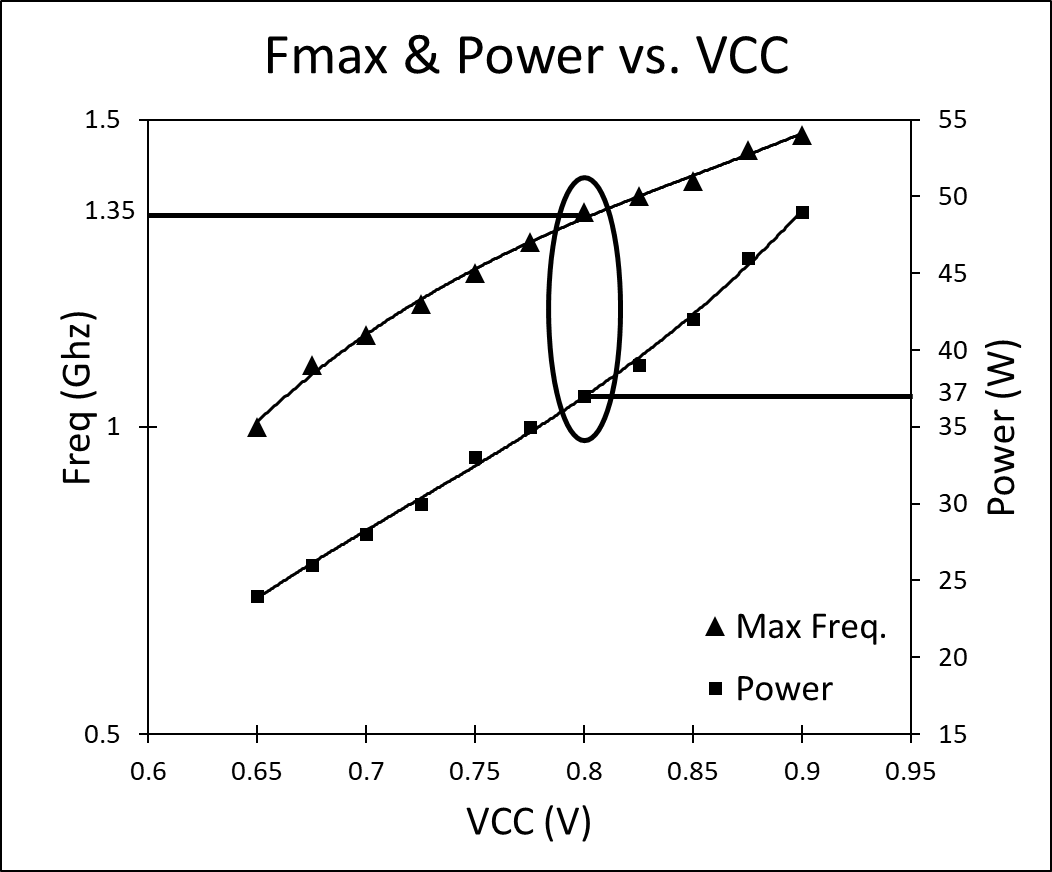}
	\includegraphics[width=.49\linewidth]{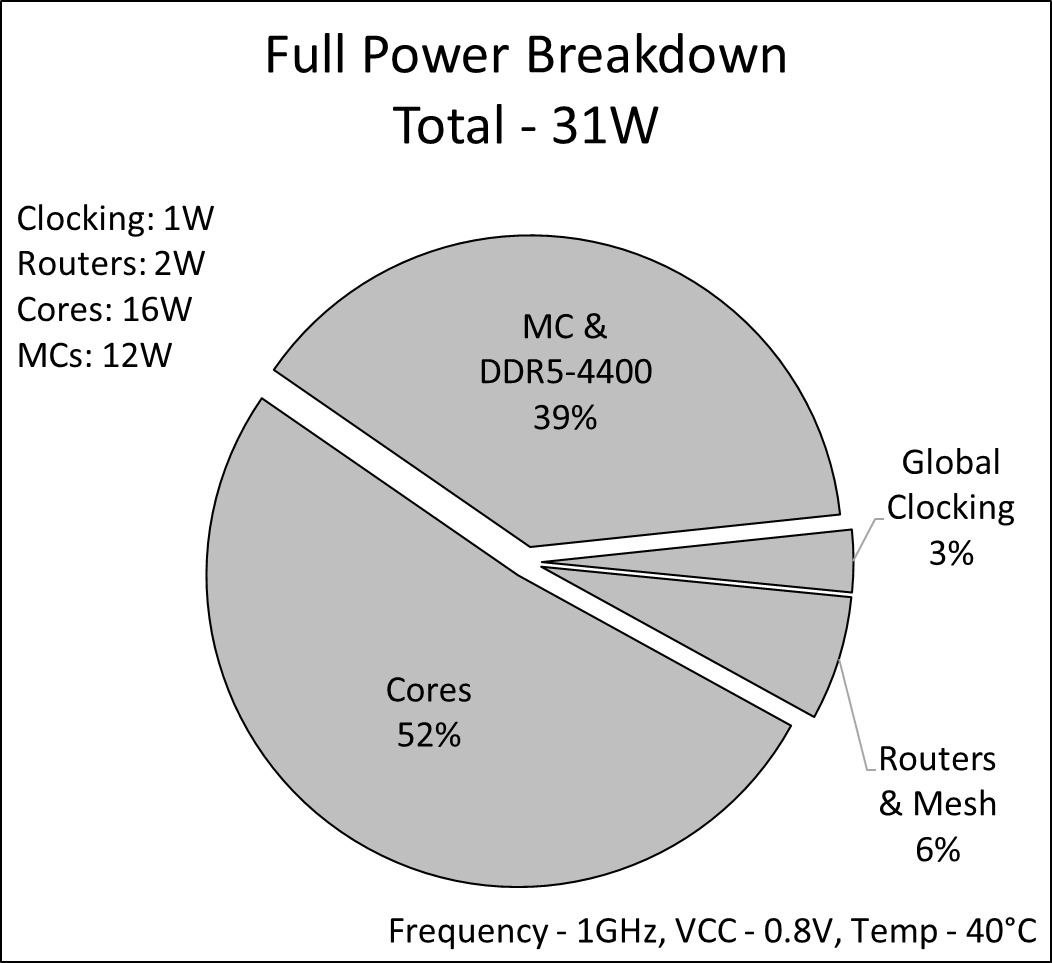}
	\caption{Maximum frequency (Fmax) and power versus VCC, and measured full-chip power breakdowns}
	\label{fig:isscc_fmax_power}
\end{figure}

\paragraph*{Power consumption}
When operating under typical conditions, 0.8~V and 1~GHz,
power consumption for a single PIUMA A0 die is measured to be 31~W at 40°C.
Maximum frequencies versus supply voltage (VCC), as well as a breakdown of power consumption by IP block,
are plotted in \cref{fig:isscc_fmax_power}.

\begin{figure}[t]
	\centering
	\includegraphics[width=\linewidth]{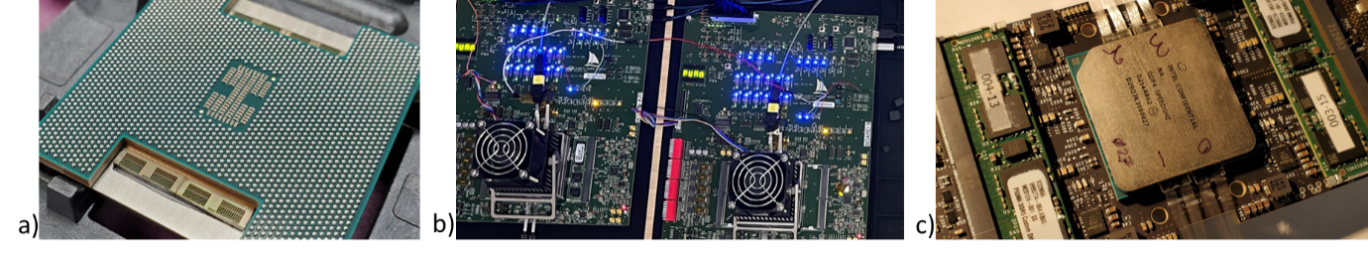}
	\caption{(a) PIUMA socket with 32 optical links; (b) two-socket debug setup; (c) socket mezzanine with DRAM and optical fibers}
	\label{fig:hotchips_socket}
\end{figure}

\begin{figure}[t]
	\centering
	\includegraphics[width=\linewidth]{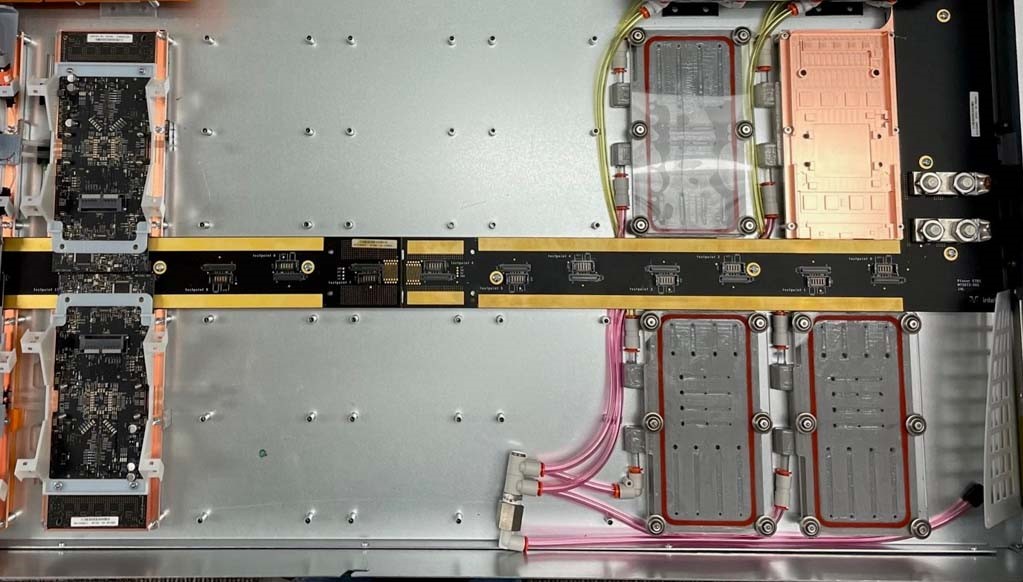}
	\caption{Partial buildout of the 16-chip PIUMA node}
	\label{fig:piuma_16node}
\end{figure}

\paragraph*{Integration}
The PIUMA prototype silicon was packaged together with the EMIB bridges and optical transceivers, and optical fibers
were attached to connect multiple sockets, see \cref{fig:hotchips_socket}.
Measured silicon demonstrates a best-case latency of $<$46ns per connection mesh-stop to mesh-stop through the optical fabric.
The largest hardware built was one PIUMA {\em node}, see \cref{fig:piuma_16node} for a picture taken during the buildout.
The node contains one x86 Xeon host processor connected via PCIe to 16 of PIUMA's A0 chips.
The PIUMA chips are fully connected via optical fiber using the lowest level of the HyperX topology and a bisectional bandwidth of 16~TB/s.

\subsection{Software bring-up}

The software stack was fully enabled in the last phase of the A0
power-on, running complete workloads on the PIUMA bring-up
boards. Because of the close co-design process, the software stack was
able to be hit the ground running. The period between tape-out and
power-on was used to focus on preparing the software
stack for execution on the A0 platform. This involved removing
simulation tricks, such as magic instructions, working around known
errata and completing the host-side development framework. This quick
software bring-up proved to be extremely valuable, as it enabled
finding hardware and software faults that arose because of rare and
complex interactions.

\subsection{Learnings}

While having a full performance simulation setup available early in
the design process was very valuable and allowed for co-design of
hardware and software, scheduling constraints meant that the A0 prototype
does not always match the architecture as designed using simulation.
For instance, to make the tape-out deadline some architectural features
and performance optimizations could not be implemented in time,
while components provided by external vendors would be late requiring
us to fall back to a previous generation with lower performance.
This means that while the original PIUMA design was very well balanced,
the actual hardware has some constraints that could have been designed
around---had we had perfect foresight of the program schedule and all
of its dependencies.

A certain class of hardware errors were found that only occur under
load, for instance, states triggered by buffer
back-pressure. These would typically only occur during complex
workload execution scenarios on the actual hardware platform. We found
that FPGA emulation or functional simulation platforms would initially
not produce these execution states, mainly because of the relative
speed difference between what is inside and outside the
simulation. For example, hardware DDR memory or a Xeon host PCIe interface
run at native speeds, so an FPGA-emulated PIUMA core running at
a few tens of MHz is unable to saturate these.
An improved validation methodology could help catch these classes of errors
earlier.

The early software bring-up on the A0 platform did present additional
challenges. Enabling a bespoke software stack on an A0 stepping of a
novel hardware architecture creates a combination of unknowns;
neither the hardware nor any software layers can be fully
trusted. Errors arising during full-stack workload runs can be
difficult and time consuming to reproduce and root cause. In such an
environment, the software stack benefits from having several
intermediate layers that can be sounded out incrementally. During the
A0 software bring-up, several new test suites were created with
increasing level of abstraction and complexity. This enabled us to test
incrementally, build trust in each layer and catch complex errors in
a simpler environment.

\section{Performance analysis}
\label{sec:results}

%
During the execution of the DARPA HIVE program, we built three testing environments (simulation,
FPGA emulation and A0 silicon) as discussed in Section~\ref{sec:codesign}. We evaluated the
functional and performance characteristics of graph applications from the DARPA HIVE
use-cases and sparse kernels that are common in these applications. The applications and the
kernels were developed using the PIUMA SDK discussed in Section~\ref{sec:software}.
In this section, we provide a summary of achieved results on PIUMA A0 silicon.
For simulation results, we refer to \cite{eyerman2020ppfiuudss} which introduces
the multi-node simulation model and shows how PIUMA performance scales to thousands of nodes.
For consistency, we decided to focus this paper on
hardware measurements only and exclude early simulation results since both the hardware
design and some of the application codes have changed---unfortunately, due to the project
ending and critical staff taking on other responsibilities, we could not rerun either
the simulations nor obtain additional hardware measurements.

During the bring-up of the A0 silicon we used a lightweight simplified runtime,
with partial support of the PIUMA software interface, to assess the system's functionality
and performance. Using a simplified runtime proved useful to quickly prototype workarounds
for hardware issues that were being discovered and fixed. We analyzed the remote and local
access latency, bandwidth to memory and scratchpad, and bandwidth achievable using DMAs
using appropriate microkernels (\eg STREAM benchmark and pointer chasing). 
Eventually we found that it was possible to port applications written for the full PIUMA SDK
to the simplified runtime with minor changes, at the cost of not being able to use more
advanced hardware functionality. The following results were collected using the simplified runtime.

\begin{table*}[t]
  \centering
  \caption{DARPA workloads, their characteristics, and inputs used to run the workloads}
  \resizebox{\textwidth}{!}{
    \begin{tabular}{p{8.2em}p{34em}p{13,5em}p{8em}p{8em}}
    \toprule
    \bf{Workload} & \bf{Description} & \bf{Limiting Factor on Xeon} & \bf{SPR input} & \bf{PIUMA input} \\
    \midrule
    \textbf{Random Walks} & Random walk in a RMAT graph & Latency & RMAT24 & RMAT18 \\
    \textbf{BFS} & Graph 500 (breadth first search) on RMAT graph & Bandwidth & RMAT24 & RMAT18 \\
    \textbf{SpMV} & Sparse matrix times dense vector on RMAT graph & Bandwidth & RMAT24 & RMAT18 \\
    \textbf{Sort} & Sorting int64 input & Bandwidth & 4.3~G & 60~K \\
    \textbf{Hash Tables} & Hash table lookup & Latency & 3.6~G & 1~M \\
    \textbf{SpGEMM} & Sparse matrix times sparse matrix on RMAT graph & Bandwidth & RMAT18 & RMAT14 \\
    \textbf{Louvain} & Louvain community detection algorithm & Bandwidth & RMAT24 & Pokec \\
    \textbf{Graph Search} & Random walk with more involved computation to select the next neigbor to visit & Latency & RMAT24 & RMAT18 \\
    \textbf{Sinkhorn} & Word movers distance \cite{DBLP:journals/corr/abs-2107-06433}, sparse selection of dense dense matrix multiplication & Bandwidth & RMAT20 & DBPEDIA-100M \\
    \textbf{Bayesian Inference} & Several kernels, Math functions, frequent barriers, Monte Carlo & Mostly Bandwidth & J1M   & J100K  \\
    \textbf{Topic Modeling} & Latent Dirichlet Allocation, Expectation Maximization, fitting posterior probabilities & Latency, limited parallelism & D8M8  & D8M8 \\
    \bottomrule
    \end{tabular}%
    }
  \label{tab:DARPAworkloads}%
\end{table*}

\begin{figure}[t]
	\centering
	\includegraphics[width=0.49\linewidth]{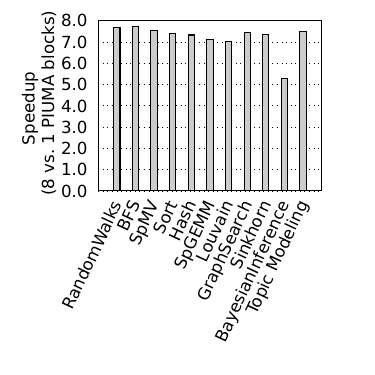}
	\includegraphics[width=0.49\linewidth]{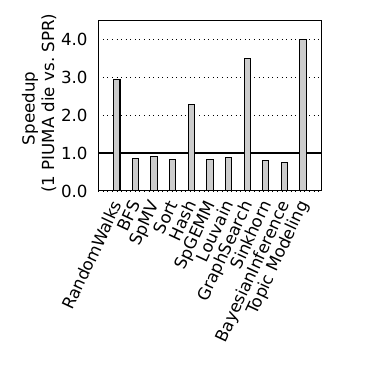}
	\caption{(a) Strong scaling of DARPA workloads on A0 silicon (b) Bandwidth-normalized comparison between Intel Xeon and PIUMA}
	\label{fig:ScalingDARPA11}
\end{figure}

We show the performance analysis of 11 DARPA workloads (see \cref{tab:DARPAworkloads}) run on A0 silicon. 
Due to hardware bugs in the memory subsystem under load we had to disable caching of modified data
to ensure reliable operation---with caching enabled, performance improved Louvain 2.45$\times$,
Bayesian Inference 1.16$\times$, and Topic Modeling 3.87$\times$ over the numbers presented below at the expense of only some runs completing successfully.
After disabling caches, we ran each workload 800 times and noticed no hardware exceptions.
\cref{fig:ScalingDARPA11}(a) shows the strong scaling of the 11 DARPA workloads on the A0 silicon.
The \(y\) axis shows the speedup obtained using 8 vs. 1 PIUMA blocks.
The GeoMean of the speedup in this case is 7.2$\times$---a 90\% scaling efficiency.

We also compared the performance of these 11 workloads with that of a run on a single-socket 60-core Intel Xeon Platinum 8490H Sapphire Rapids (SPR) \cite{nassif2022sapphire}
machine running two threads per core at 2.5~GHz, with 115~MB of last-level cache (LLC) capacity and 244~GB/s main memory (DDR5) bandwidth.
The inputs are large enough to ensure the data does not fit in the LLC on SPR (PIUMA performance is insensitive to graph size
since---apart from code and local variables---nothing is cached anyway).
Performance was measured as throughput,
in number of operations per second using a metric defined by each workload (\eg edge traversals per second).
To make the comparison fair, we normalize the results by available bandwidth (244~GB/s for SPR vs. 35~GB/s for PIUMA).
The power consumption has roughly the same ratio (350~W for one socket of SPR vs. 37~W for PIUMA) so to a first order,
the energy efficiency of PIUMA relative to SPR is similar to its relative performance.

These workloads were not optimized with specialized SPR features (\eg AMX, DSA, etc.),
since most of them are either latency-bound or bandwidth-bound. However, we used the latest compilers and
best-known compiler optimization flags to compile the code on SPR.
At the same time, PIUMA performance was hampered by the fact that we had to disable caches.
\cref{fig:ScalingDARPA11}(b) shows the performance comparison between SPR and the A0 silicon.
For the bandwidth-bound workloads, we can see PIUMA is able to keep up with SPR.
PIUMA's worst result is for Bayesian Inference, where the input data we had
was only slightly larger than the SPR LLC and thus, did benefit from caching.
For the latency bound and mixed resource bound workloads, we see a significant speedup of PIUMA over SPR,
up to 4.0$\times$ for Topic Modeling.
This occurs even though SPR clocks at a 2.5$\times$ higher frequency, and applications may benefit from
some automatic vectorizations, caching, and prefetching. Still, due to instruction dependencies and a limited
number of hardware threads, SPR is not able to saturate its memory bandwidth for these workloads.
This shows PIUMA's strenght of employing massive thread-level parallelism to deal with latency-bound applications
and hide latency for random accesses better than a standard Xeon.
In summary, the performance we observed on the A0 silicon is aligned with our expectations of an early silicon prototype.
Performance is competitive with the latest Intel Xeon server processors, and should handily outperform it on a per-node basis
in both absolute performance and energy efficiency assuming we fixed the bugs in PIUMA A0's memory subsystem.

\section{Conclusions}

PIUMA is a graph analysis oriented architecture developed by Intel in response to the DARPA HIVE project.
Based on the observation that graph workloads are dominated by irregular sparse accesses, it features many highly-threaded simple cores to hide the latency of remote memory accesses.
Combined with small access granularity to memory and network, and powerful offload engines, PIUMA outperforms current high-end processors for typical graph workloads.
Furthermore, it is designed to scale out efficiently thanks to the high bandwidth network and shared address space, increasing the performance gap with current multi-node computers, which perform poorly on distributed graph applications.

We built an A0 test chip, powered it on in the lab and were able to run workloads on it.
The effective hardware/software co-design process of PIUMA guaranteed highly optimized hardware, and ensured that system and development tools were available by the time we had silicon in the lab.
Despite finding hardware bugs that requiring software workarounds, 
we were able to validate our performance and energy efficiency targets that make PIUMA a highly capable graph analytics platform.

\section*{Acknowledgements}
This research was developed with funding from the Defense Advanced Research Projects Agency (DARPA).
The views, opinions and/or findings expressed are those of the author and should not be interpreted as representing
the official views or policies of the Department of Defense or the U.S. Government.

Distribution Statement A---Approved for Public Release, Distribution Unlimited


\balance

\bibliographystyle{IEEEtranS}
\bibliography{refs}

\end{document}